\documentclass[seceq]{ptptex}
\usepackage[dvips]{graphicx}
\usepackage{bm,latexsym,amsmath,amssymb,amsfonts,mathrsfs}
\usepackage{color}
\input{colordvi.tex}
\newcommand*{\D}{{\rm d}}
\newcommand*{\mpl}{M_{\rm Pl}}
\newcommand*{\calk}{{\cal K}}
\begin{document}

\markboth{
T. Kobayashi, M. Yamaguchi, and J. Yokoyama%
}{
Generalized G-inflation%
}

\title{Generalized G-inflation:
Inflation with the most general second-order field equations}

\author{
Tsutomu \textsc{Kobayashi}$^{a,b}$,
Masahide \textsc{Yamaguchi}$^c$, and
Jun'ichi \textsc{Yokoyama}$^{d,e}$%
}

\inst{
$^a$Hakubi Center, Kyoto University, Kyoto 606-8302, Japan\\
$^b$Department of Physics, Kyoto University, Kyoto 606-8502, Japan\\
$^c$Department of Physics, Tokyo Institute of Technology, Tokyo 152-8551, Japan\\
$^d$Research Center for the Early Universe (RESCEU), Graduate School of Science,
The University of Tokyo, Tokyo 113-0033, Japan\\
$^e$Institute for the Physics and Mathematics of the Universe (IPMU),\\
The University of Tokyo, Kashiwa, Chiba, 277-8568, Japan
}

\abst{
We study generalized Galileons as a framework to develop the most
general single-field inflation models ever, Generalized G-inflation,
containing yet further generalization of G-inflation, as well as
previous examples such as k-inflation, extended inflation, and new Higgs
inflation as special cases.  We investigate the background and
perturbation evolution in this model, calculating the most general
quadratic actions for tensor and scalar cosmological perturbations to
give the stability criteria and the power spectra of primordial
fluctuations.
It is pointed out in the Appendix
that the Horndeski theory and the generalized Galileons are equivalent.
In particular, even the non-minimal coupling to the Gauss-Bonnet term is
included in the generalized Galileons in a non-trivial manner.
}

\maketitle

\section{Introduction}

Scalar fields play important roles in cosmology.  On the one hand, inflation
in the early Universe is now becoming a part of 
standard cosmology that is driven by a scalar
field called the inflaton~\cite{inflation,r2,slowroll}.  
The conventional inflaton
action consists of a canonical kinetic term and a sufficiently flat
potential \cite{slowroll}. [See Ref.~\citen{Maz} for the latest review.]
Non-canonical kinetic terms~\cite{kinflation} also arise naturally in some
particle physics models of inflation such as Dirac-Born-Infeld
inflation~\cite{dbi}.  

On the other hand, it is strongly suggested that the present
Universe is dominated by mysterious dark energy, and its identity might
be a dynamical scalar field~\cite{essence}.  In relation to the present
accelerated expansion, modified gravity theories have been studied
extensively, and in such theories, an extra gravitational degree of
freedom can often be equivalently described by a scalar field coupled
non-minimally to gravity or matter.  In the decoupling limit of the
Dvali-Gabadadze-Porrati brane model~\cite{dgp}, the scalar field has a
non-linear derivative self-interaction~\cite{declim}, which was later
generalized to Galileons~\cite{Galileon1} with a number of applications
to various contexts in cosmology~\cite{gcos, tsujikaw, KGB, G-inf, kamada, nonG,
KGB2, KimuraYamamoto, genesis, Mizuno:2010ag, galinf, Creminelli,
Naruko}.  
Thus, in recent years, there have been growing interests in
scalar field theories beyond the canonical one.  

The most attractive feature of  higher derivative theories possessing
the Galilean invariance
$\partial_\mu\phi\to\partial_\mu\phi + b_\mu$
is that field equations derived from such a
theory contain derivatives only up to second order~\cite{Galileon1}, 
so that it can easily avoid ghosts.
Unfortunately, however, this desired feature ceases to exist once
the background spacetime is curved~\cite{CovariantGalileon}.

To preserve the second-order nature of field equations,
the ``covariantization'' of the Galileon has been proposed
by Deffayet et al.~\cite{CovariantGalileon} where
the theory is no longer Galilean invariant.
This line of analysis has been further pursued recently to yield
a more generic class of higher-derivative theories that result in
second-order field equations~\cite{GeneralizedGalileon}.
In this theory, the Galilean invariance is absent even in the flat spacetime
limit.

The purpose of this paper is to provide a comprehensive and
thorough study
of the most general non-canonical and non-minimally coupled
single-field inflation models yielding
 second-order field equations
making use of Ref.~\citen{GeneralizedGalileon}, which is the
most general extension of the Galileons but is no longer
based on a symmetry argument.
It would be nice if one could develop a new class of viable inflation
models fully respecting the Galileon symmetry, $\phi\to
\phi+b_\mu x^\mu+c$, as was attempted in Refs.~\citen{galinf,Creminelli}, 
but such a symmetry must be actually broken to construct a
phenomenologically viable inflation model, namely, to terminate 
inflation and reheat the Universe.
Thus, our strategy here is more similar to the G-inflation
model~\cite{G-inf} and is indeed the most general extension of it.
Hence, one may call the model presented here as
{\em Generalized G-inflation} or {\em G$^2$-inflation}.

Special cases of the generalized Galileons can be
derived from
a relativistic probe brane embedded in a five-dimensional
bulk~\cite{deRhamTolley, Trodden, Burrage:2011bt, petel},
and hence, they are possibly related to fundamental theory and particle physics.
The scalar field theories we are going to study thus
include
not only all the previous examples
considered in the context of single-field inflation,
but also recent developments and their further generalization.
We clarify the generic behavior of the inflationary background and
investigate the nature of primordial tensor and scalar perturbations at linear order.
Given a specific model, our formulas are helpful
to determine the evolution of cosmological perturbations and
its observational consequences.

This paper is organized as follows.
In the next section, we define the scalar-field theories that we consider.
We then provide the background cosmological equations in \S3,
and explore two possible inflationary mechanisms.
In \S4, cosmological perturbations are considered
and the quadratic actions for tensor and scalar perturbations are
computed, which are used to give stability conditions and
evaluate the primordial power spectra.
Our conclusion is drawn in \S5.

\section{Generalized Higher-order Galileons and Kinetic Gravity Braiding}

Galileons~\cite{Galileon1} and their covariant 
extension~\cite{CovariantGalileon}
have been further  generalized recently to yield
the most general scalar field theories having
second-order field equations~\cite{GeneralizedGalileon}.
The first two terms of the generalized Lagrangian 
corresponding to $(\partial\phi)^2$ and 
$(\partial\phi)^2\Box\phi$ in the original theory
are given by~\cite{KGB, G-inf}
\begin{eqnarray}
{\cal L}_2&=&K(\phi, X),
\\
{\cal L}_3&=&-G_3(\phi, X)\Box\phi,
\end{eqnarray}
where $K$ and $G_3$ are generic functions of $\phi$ and $X:=-\partial_\mu\phi \partial^\mu\phi/2$.
Similarly, higher-order Galileons can be generalized to give~\cite{GeneralizedGalileon}
\begin{eqnarray}
{\cal L}_4&=&G_{4}(\phi, X)R+G_{4X}\left[
\left(\Box\phi\right)^2-\left(\nabla_\mu\nabla_\nu\phi\right)^2
\right],
\\
{\cal L}_5&=&G_5(\phi, X) G_{\mu\nu}\nabla^\mu\nabla^\nu\phi
-\frac{G_{5X}}{6}\Bigl[
\left(\Box\phi\right)^3
-3\left(\Box\phi\right)\left(\nabla_\mu\nabla_\nu\phi\right)^2
+2\left(\nabla_\mu\nabla_\nu\phi\right)^3
\Bigr],\,\,\,\,\,\,
\end{eqnarray}
where $R$ is the Ricci tensor, $G_{\mu\nu}$ is the Einstein tensor,
$(\nabla_\mu\nabla_\nu\phi)^2=\nabla_\mu\nabla_\nu\phi\nabla^\mu\nabla^\nu\phi$,
$(\nabla_\mu\nabla_\nu\phi)^3=\nabla_\mu\nabla_\nu\phi\nabla^\nu\nabla^\lambda\phi\nabla_\lambda\nabla^\mu\phi$,
and $G_{iX}=\partial G_i/\partial X$.
Setting $G_3=X$, $G_4=X^2$, and $G_5=X^2$, the above Lagrangians reproduce
the covariant Galileons introduced in Ref.\ \citen{CovariantGalileon}.
The non-minimal couplings to gravity in ${\cal L}_4$ and ${\cal L}_5$
are necessary to eliminate higher derivatives
that would otherwise appear in the field equations.
Note that we do not need a separate gravitational Lagrangian
other than ${\cal L}_4$;
for $G_4=\mpl^2/2$, ${\cal L}_4$ reduces to the Einstein-Hilbert term.
We also obtain a non-minimal coupling of the form $f(\phi) R$
from ${\cal L}_4$ by taking $G_4=f(\phi)$.
The non-standard kinetic term $G^{\mu\nu}\partial_\mu\phi\partial_\nu\phi$
that is considered, such as in Ref.~\citen{Germani},
turns out to be a special case $G_5\propto \phi$
of ${\cal L}_5$ after integration by parts.
Equation~(24) of Ref.~\citen{VanAcoleyen:2011mj}, which is obtained from
a Kaluza-Klein compactification of higher-dimensional Lovelock gravity,
turns out to be equivalent to ${\cal L}_5$ with $G_5=-3X/2$.

We thus consider a gravity + scalar system described by the action
\begin{eqnarray}
S=\sum_{i=2}^{5}\int\D^4x\sqrt{-g}{\cal L}_i, \label{5}
\end{eqnarray}
which is the most general single scalar theory resulting in equations of
motion containing derivatives up to second order.  This action
contains only four independent arbitrary functions of $\phi$ and $X$.
This theory represents a general class of single-field inflation,
including models that have not been studied so far, as well as almost
all the previously known models such as potential-driven slow-roll
inflation \cite{slowroll}, k-inflation~\cite{kinflation}, extended
inflation~\cite{extinf}, and even new Higgs inflation~\cite{Germani}
as special cases.\footnote{Even the curvature-square inflation model as well
as more general $f(R)$ inflation~\cite{r2,r2m}, which do not
contain any scalar field and result in fourth-order equations of motion,
can be recast in the present form by defining a new field as $\phi=\D f/\D R$.}




Note in passing that
\begin{eqnarray*}
XR+(\Box\phi)^2-(\nabla_\mu\nabla_\nu\phi)^2
&=&G_{\mu\nu }\nabla^\mu\phi\nabla^\nu\phi
+{\rm total~derivative}
\nonumber\\
&=&-\phi G_{\mu\nu }\nabla^\mu\nabla^\nu\phi
+{\rm total~derivative},
\end{eqnarray*}
which implies that ${\cal L}_4$ with $G_4=X$ and
${\cal L}_5$ with $G_5= -\phi$ are equivalent.
Similarly, ${\cal L}_3$ with $G_3=f(\phi)$ and
${\cal L}_2$ with $K=-2X f_{\phi}$ coincide up to total derivative.
These facts can be used to check the calculations.

\section{Background equations}

Let us derive the equations of motion describing the background
evolution from (\ref{5}).
The easiest way is to substitute $\phi=\phi(t)$ and
the metric $\D s^2 =-N^2(t)\D t^2+a^2(t)\D \mathbf{x}^2$
to the action. Variation with respect to $N(t)$ gives the constraint equation
corresponding to the Friedmann equation, which can be written as
\begin{eqnarray}
\sum_{i=2}^5{\cal E}_i = 0, \label{E}
\end{eqnarray}
where
\begin{eqnarray}
{\cal E}_2&=&2XK_X-K,
\\
{\cal E}_3&=&6X\dot\phi HG_{3X}-2XG_{3\phi},
\\
{\cal E}_4&=&-6H^2G_4+24H^2X(G_{4X}+XG_{4XX})
-12HX\dot\phi G_{4\phi X}-6H\dot\phi G_{4\phi },
\\
{\cal E}_5&=&2H^3X\dot\phi\left(5G_{5X}+2XG_{5XX}\right)
-6H^2X\left(3G_{5\phi}+2XG_{5\phi X}\right).
\end{eqnarray}
The above quantities
contain derivatives of the metric and the scalar field up to first order.

Variation with respect to $a(t)$ yields the evolution equation,
\begin{eqnarray}
\sum_{i=2}^5{\cal P}_i=0,  \label{P}
\end{eqnarray}
where
\begin{eqnarray}
{\cal P}_2&=&K,
\\
{\cal P}_3&=&-2X\left(G_{3\phi}+\ddot\phi G_{3X} \right) ,
\\
{\cal P}_4&=&2\left(3H^2+2\dot H\right) G_4
-12 H^2 XG_{4X}-4H\dot X G_{4X}
-8\dot HXG_{4X}-8HX\dot X G_{4XX}
\nonumber\\&&
+2\left(\ddot\phi+2H\dot\phi\right) G_{4\phi}
+4XG_{4\phi\phi}
+4X\left(\ddot\phi-2H\dot\phi\right) G_{4\phi X},
\\
{\cal P}_5&=&-2X\left(2H^3\dot\phi+2H\dot H\dot\phi+3H^2\ddot\phi\right)G_{5X}
-4H^2X^2\ddot\phi G_{5XX}
\nonumber\\&&
+4HX\left(\dot X-HX\right)G_{5\phi X}
+2\left[2\left(HX\right){\bf \dot{}}+3H^2X\right]G_{5\phi}
+4HX\dot\phi G_{5\phi\phi}.
\end{eqnarray}

The background quantities ${\cal E}_i$ and ${\cal P}_i$ are defined in
an analogous way in which the energy density and the isotropic pressure
of a usual scalar field are defined.  In the present case, however, the
distinction between the gravitational and scalar-field portions of the
Lagrangian is ambiguous, and hence, in that sense, the gravitational
contribution is included in the above expressions. Indeed, one can see
that, for $G_4=\mpl^2/2$, ${\cal E}_4$ and ${\cal P}_4$ reduce to the
``minus'' of the Einstein tensor $G_\mu^{\;\nu}$: ${\cal
E}_4=-3\mpl^2H^2$ and ${\cal P}_4=\mpl^2(3H^2+2\dot H)$.

%

Variation with respect to $\phi(t)$ gives the scalar-field equation of motion,
\begin{eqnarray}
\frac{1}{a^3}\frac{\D}{\D t}\left(a^3 J\right) =P_\phi,\label{phieom}
\end{eqnarray}
where
\begin{eqnarray}
J&=&\dot\phi K_X+6HXG_{3X}-2\dot\phi G_{3\phi}
+6H^2\dot\phi\left(G_{4X}+2XG_{4XX}\right)-12HXG_{4\phi X}
\nonumber\\&&
+2H^3X\left(3G_{5X}+2XG_{5XX}\right)
-6H^2\dot\phi\left(G_{5\phi}+XG_{5\phi X}\right), 
\end{eqnarray}
and
\begin{eqnarray}
P_\phi &=&
K_\phi-2X\left(G_{3\phi\phi}+\ddot\phi G_{3\phi X}\right)
+6\left(2H^2+\dot H\right)G_{4\phi}
+6H\left(\dot X+2HX\right)G_{4\phi X}
\nonumber\\&&
-6H^2XG_{5\phi\phi}+2H^3X\dot\phi G_{5\phi X}.
\end{eqnarray}
Not all of the equations~(\ref{E}),~(\ref{P}), and~(\ref{phieom}) are mutually
independent: Eq.~(\ref{phieom}) [(\ref{P})] can be derived from
Eqs.~(\ref{P}) [(\ref{phieom})] and~(\ref{E}).

Below, we present two extreme cases of inflation, one driven purely
kinetically, which is an extension of the original G-inflation
\cite{G-inf} corresponding to the case with $G_4=\mpl^2/2$ and $G_5=0$,
the other driven by a scalar potential as an extension of Higgs
G-inflation \cite{kamada}.

\subsection{Kinetically driven G-inflation}

Let us start with a shift-symmetric model, $\phi\to\phi+c$.
This in particular implies that $\phi$ does not have any potential.
In this case, the field equations are
\begin{eqnarray}
\sum_{i=2}^5{\cal E}_i
&=&\dot\phi J-K-6H^2(G_4-2XG_{4X})
+4H^3X\dot\phi G_{5X}=0,
\\
\sum_{i=2}^5({\cal E}_i+{\cal P}_i)
&=&\dot\phi J-2X\ddot\phi G_{3X}
+2\frac{\D}{\D t}\Bigl[
2H(G_4-2XG_{4X})
-H^2X\dot\phi G_{5X}
\Bigr]
\nonumber\\
&=&0,
\\
\frac{\D}{\D t}(a^3 J)&=&0.\label{shiftsymmeom}
\end{eqnarray}
From Eq.~(\ref{shiftsymmeom}),
one immediately finds that $J\propto a^{-3}\to 0$.
Shift-symmetric models thus have an attractor, $J=0$,
along which $H=$const, $\dot\phi=$const, satisfying
\begin{eqnarray}
\dot\phi K_X+6H XG_{3X}
+6H^2\dot\phi\left(G_{4X}+2XG_{4XX}\right)&&
\nonumber\\
+2H^3X\left(3G_{5X}+2XG_{5XX}\right)&=&0,\;\;\;\;\;\label{kg1}
\\
K+6H^2(G_4-2XG_{4X})-4H^3X\dot\phi G_{5X}&=&0.\label{kg2}
\end{eqnarray}
Provided that Eqs.~(\ref{kg1}) and~(\ref{kg2}) have a non-trivial root,
$H\neq 0$, $\dot\phi\neq 0$,
we obtain inflation driven by $\phi$'s kinetic energy.
This is the generalization of kinetically driven G-inflation~\cite{G-inf}
and kinetic gravity braiding~\cite{KGB}.

Although the shift-symmetric Lagrangian can nicely accommodate a de
Sitter solution as an attractor,
the shift symmetry must be broken in some region in the field space
to end inflation and to reheat the Universe.
Following the arguments in Refs.~\cite{kinflation, G-inf},
a graceful exit from kinetically driven inflation is
possible  with gravitational
reheating~\cite{Ford}. However,
a detailed analysis of the reheating stage after kinetically driven inflation
is beyond the scope of the present paper.

\subsection{Potential-driven slow-roll G-inflation}\label{slowroll}

Suppose that the functions in the Lagrangian can be expanded in terms of
$X$ as 
\begin{eqnarray}
K(\phi, X)&=&-V(\phi)+\calk(\phi)X+\cdots,
\\
G_i(\phi, X)&=&g_i(\phi)+h_i(\phi)X+\cdots,
\end{eqnarray}
and consider the case in which the inflaton field value $\phi(t)$ changes very slowly.
In this case, the potential term manifestly breaks the shift symmetry, and thereby,
the model is capable of a graceful exit from inflation~\cite{Kawasaki:2000yn}.
Neglecting all the terms multiplied by $\dot\phi$ in the gravitational field equations,
we obtain
\begin{eqnarray}
\sum_{i=2}^5{\cal P}_i\simeq
-\sum_{i=2}^5{\cal E}_i\simeq -V(\phi)+6g_4(\phi)H^2,
\end{eqnarray}
where we have assumed
\begin{eqnarray}
|\dot H|\ll H^2
~~{\rm and}~~
|\ddot\phi|\ll| H\dot\phi|.
\end{eqnarray}
We may thus have slow-roll inflation with
\begin{eqnarray}
H^2\simeq \frac{V}{6g_4}.\label{frpot}
\end{eqnarray}

During slow-roll, we approximate
\begin{eqnarray}
|\dot J|\ll |HJ|,
\quad
|\dot g_i|\ll |H g_i|,
\quad
|\dot h_i|\ll |H h_i|.
\end{eqnarray}
Under the above approximation, we have the slow-roll
equation of motion for $\phi$,
\begin{eqnarray}
3H J \simeq -V_\phi  + 12 H ^2g_{4\phi},
\end{eqnarray}
with
\begin{eqnarray}
J \simeq \calk\dot\phi - 2g_{3\phi}\dot\phi
+ 6\left(Hh_3 X+H^2h_4\dot\phi-H^2g_{5\phi}\dot\phi+
H^3h_5X\right).
\label{slowrollJ}
\end{eqnarray}
Which term is dominant in Eq.~(\ref{slowrollJ})
depends on the magnitude of the coefficients $h_i(\phi)$ of $X$.
Note here that we can set $g_3=0$
and $g_5=0$ without loss of generality, because $g_{3\phi}$ can be
absorbed into the redefinition of $\calk$ and $g_{5\phi}$ into $h_4$,
that is, $\calk-2g_{3\phi} \rightarrow \calk,~~
h_4-g_{5\phi} \rightarrow h_4$.

Equations~(\ref{frpot}) and~(\ref{slowrollJ}) imply that
the dominant contribution to the inflationary Hubble parameter is the potential
$V$ in ${\cal L}_2$,
while any of the terms in Eq.~(\ref{slowrollJ})
can participate to determine the actual dynamics of the scalar field.
Therefore, the slow-roll parameters expressed in terms of the potential may look very different
from the standard ones in general.
This is the generalization of the Higgs G-inflation~\cite{kamada} 
(see also Ref.~\citen{tsuji}).

\section{Quadratic actions for tensor and scalar perturbations}

In this section, our goal is to compute quadratic actions
for tensor and scalar cosmological perturbations in Generalized
G-inflation.
We use the unitary gauge in which $\phi=\phi(t)$ and begin with
writing the perturbed metric as
\begin{eqnarray}
\D s^2=-N^2\D t^2+\gamma_{ij}\left(\D x^i +N^i\D t\right)\left(\D x^j +N^j\D t\right),
\end{eqnarray}
where
\begin{eqnarray}
N=1+\alpha,\quad
N_i=\partial_i\beta,\quad
\gamma_{ij} =a^2(t)e^{2\zeta}\left(\delta_{ij}+h_{ij}+\frac{1}{2}h_{ik}h_{kj}\right).
\end{eqnarray}
Here, $\alpha$, $\beta$, and $\zeta$ are scalar perturbations and
$h_{ij}$ is a tensor perturbation satisfying $h_{ii}=0=h_{ij,j}$.
With the above definition of the perturbed metric,
$\sqrt{-g}$ does not contain $h_{ij}$ up to second order,
and the coefficients of $\zeta^2$ and $\alpha\zeta$ vanish,
thanks to the background equations.

\subsection{Tensor perturbations}

The quadratic action for the tensor perturbations is found to be
\begin{eqnarray}
S_T^{(2)} =\frac{1}{8}\int\D t\D^3x\,a^3\left[
{\cal G}_T\dot h_{ij}^2-\frac{{\cal F}_T}{a^2}
(\Vec{\nabla} h_{ij})^2\right], \label{tensoraction}
\end{eqnarray}
where
\begin{eqnarray}
{\cal F}_T&:=&2\left[G_4
-X\left( \ddot\phi G_{5X}+G_{5\phi}\right)\right],
\\
{\cal G}_T&:=&2\left[G_4-2 XG_{4X}
-X\left(H\dot\phi G_{5X} -G_{5\phi}\right)\right].
\end{eqnarray}
One may notice that ${\cal G}_T$ can also be expressed as
\begin{eqnarray}
{\cal G}_T=\frac{1}{2}\sum_{i=2}^5\frac{\partial{\cal P}_i}{\partial \dot H}.
\label{comp1}
\end{eqnarray}
The squared sound speed is given by
\begin{eqnarray}
c_T^2=\frac{{\cal F}_T}{{\cal G}_T}.
\end{eqnarray}
One sees from the action~(\ref{tensoraction})
that ghost and gradient instabilities are avoided
provided that
\begin{eqnarray}
{\cal F}_T>0,\quad {\cal G}_T>0.\label{stts}
\end{eqnarray}
Note that $c_T^2$ is not necessarily unity in general cases, contrary
to the standard inflation models.

To canonically normalize the tensor perturbation, we define
\begin{eqnarray}
\D y_T := \frac{c_T}{a}\D t,
\;\;
z_T:=\frac{a}{2} \left({\cal F}_T{\cal G}_T\right)^{1/4},
\;\;
v_{ij}:= z_T h_{ij},
\end{eqnarray}
and then the quadratic action is written as
\begin{eqnarray}
S_T^{(2)} =\frac{1}{2}\!\int\D y_T\D^3 x\left[
(v_{ij}')^2- (\Vec{\nabla} v_{ij} )^2+\frac{z_T''}{z_T}v_{ij}^2
\right],\label{tens-v-action}
\end{eqnarray}
where a prime denotes differentiation with respect to $y_T$.
In terms of the Fourier wavenumber $k$,
sound horizon crossing occurs when $k^2=z_T''/z_T\sim 1/y^2_T$
for each mode.

On superhorizon scales,
the two independent solutions to the perturbation
equation that follows from the action~(\ref{tens-v-action}) are
\begin{equation}
v_{ij}\propto z_T\quad{\rm and}\quad z_T\int \frac{\D y_T}{z_T^2}.
\end{equation}
In terms of the original variables, the two independent solutions
on superhorizon scales are given by
\begin{eqnarray}
h_{ij} = {\rm const}\quad{\rm and}\quad \int^t\frac{\D t'}{a^3{\cal G}_T}.
\label{shsolgw}
\end{eqnarray}
The second solution corresponds to a decaying mode.

To evaluate the primordial power spectrum, let us assume that
$\epsilon:=-\dot H/H^2\simeq\;$const,
\begin{eqnarray}
f_{T}:=\frac{\dot{{\cal F}}_T}{H{\cal F}_T}\simeq {\rm const}
~~{\rm and}~~
g_{T}:=\frac{\dot{{\cal G}}_T}{H{\cal G}_T}\simeq {\rm const}.
\end{eqnarray}
We also define the variation parameter of the sound velocity of tensor
perturbations as
\begin{eqnarray}
s_{T}:=\frac{\dot{c}_T}{H c_T} = \frac12 \left(f_{T} - g_{T} \right).
\end{eqnarray}
Clearly, only two of the three parameters are independent. We additionally impose
conditions
\begin{eqnarray}
1-\epsilon -f_T/2+g_T/2>0, \label{T1} \\
3-\epsilon+g_T>0. \label{T2}
\end{eqnarray}
The former (equivalent to $1 > \epsilon + s_{T}$) guarantees that
the time coordinate $y_T$ runs from $-\infty$ to $0$ as the Universe expands.
The latter implies that the second solution in~(\ref{shsolgw}) indeed decays.
We see that $z_T$ can be written as
\begin{eqnarray}
z_T=\frac{{\cal F}_{T*}^{3/4}}{2{\cal G}_{T*}^{1/4}}\frac{1}{H_*(-y_{T*})}
\frac{[(-y_T)/(-y_{T*})]^{1/2-\nu_T }}{1-\epsilon-f_T/2+g_T/2},
\end{eqnarray}
where the quantities with $*$ are those evaluated at some reference time $y_T=y_{T*}$. 
The normalized mode solution to the perturbation equation is given
in terms of the Hankel function:
\begin{eqnarray}
v_{ij}=\frac{\sqrt{\pi}}{2}\sqrt{-y}H_{\nu_T}^{(1)}(-ky_T) \,{\rm e}_{ij},
\end{eqnarray}
where
\begin{eqnarray}
\nu_T:=\frac{3-\epsilon+g_T}{2-2\epsilon-f_T+g_T},
\end{eqnarray}
and ${\rm e}_{ij}$ is a polarization tensor. Notice that the
conditions (\ref{T1}) and (\ref{T2}) guarantee the positivity of
$\nu_T$.  On superhorizon scales, $-k y_T\ll 1$, we obtain
\begin{eqnarray}
k^{3/2}h_{ij}\approx 2^{\nu_T -2}\frac{\Gamma(\nu_T)}{\Gamma(3/2)}
\frac{(-y_T)^{1/2-\nu_T}}{z_T}k^{3/2-\nu_T }
{\rm e}_{ij}. 
\end{eqnarray}
Thus, we find the power spectrum of the primordial tensor perturbation:
\begin{eqnarray}
{\cal P}_T=8\gamma_T\left.
\frac{{\cal G}_{T}^{1/2}}{{\cal F}_{T}^{3/2}}
\frac{H^2}{4\pi^2}\right|_{-ky_T=1},
\end{eqnarray}
where $\gamma_T=2^{2\nu_T -3}|\Gamma(\nu_T)/\Gamma(3/2)|^2(1-\epsilon -f_T/2+g_T/2)$.
The tensor spectral tilt is given by
\begin{eqnarray}
n_T=3-2\nu_T .
\end{eqnarray}

Contrary to the predictions of the conventional inflation models,
the blue spectrum $n_T > 0$ can be obtained if the following condition
is satisfied,
\begin{eqnarray}
 4 \epsilon + 3 f_T - g_T < 0.
\end{eqnarray}
This condition is easily compatible with the conditions (\ref{T1}) and
(\ref{T2}).  Thus, positive and large $g_T$ compared with $\epsilon$ and
$f_T$ can lead to a blue spectrum of tensor perturbations. In deriving
the above formulas, we only assumed that $\epsilon$, $f_T$, and $g_T$ are
constant. These parameters may not necessarily be very small as long as
the inequalities (\ref{T1}) and (\ref{T2}) are satisfied under a sensible
background solution.



%
%
%

\subsection{Scalar perturbations}

We now focus on scalar fluctuations putting $h_{ij}=0$.
Plugging the perturbed metric into the action and
expanding it to second order, we obtain
\begin{eqnarray}
S^{(2)}_S&=&\int \D t\D^3 x a^3\biggl[
-3{\cal G}_{T}\dot\zeta^2+\frac{{\cal F}_T}{a^2}(\Vec{\nabla}\zeta)^2
+\Sigma \alpha^2
\nonumber\\&&\qquad\qquad\qquad
-2\Theta\alpha\frac{\Vec{\nabla}^2}{a^2}\beta
+2{\cal G}_T\dot \zeta\frac{\Vec{\nabla}^2}{a^2}\beta
+6\Theta \alpha\dot\zeta-2{\cal G}_T\alpha\frac{\Vec{\nabla}^2}{a^2}\zeta
\biggr],\label{scalaraction1}
\end{eqnarray}
where
\begin{eqnarray}
\Sigma&:=&XK_X+2X^2K_{XX}+12H\dot\phi XG_{3X}
\nonumber\\&&
+6H\dot\phi X^2G_{3XX}
-2XG_{3\phi}-2X^2G_{3\phi X}-6H^2G_4
\nonumber\\&&
+6\Bigl[H^2\left(7XG_{4X}+16X^2G_{4XX}+4X^3G_{4XXX}\right)
\nonumber\\&&
-H\dot\phi\left(G_{4\phi}+5XG_{4\phi X}+2X^2G_{4\phi XX}\right)
\Bigr]
\nonumber\\&&
+30H^3\dot\phi XG_{5X}+26H^3\dot\phi X^2G_{5XX}
\nonumber\\&&
+4H^3\dot\phi X^3G_{5XXX}
-6H^2X\bigl(6G_{5\phi}
+9XG_{5\phi X}+2 X^2G_{5\phi XX}\bigr),
\\
\Theta&:=&-\dot\phi XG_{3X}+
2HG_4-8HXG_{4X}
-8HX^2G_{4XX}+\dot\phi G_{4\phi}+2X\dot\phi G_{4\phi X}
\nonumber\\&&
-H^2\dot\phi\left(5XG_{5X}+2X^2G_{5XX}\right)
+2HX\left(3G_{5\phi}+2XG_{5\phi X}\right).
\end{eqnarray}
It is interesting to see that even in the most generic case,
some of the coefficients are given by ${\cal F}_T$ and ${\cal G}_T$,
{\it i.e.}, the functions characterizing the tensor perturbation, and
only two new functions show up in the scalar quadratic action.
Note that the following relations hold:
\begin{eqnarray}
\Sigma &=& X\sum_{i=2}^5\frac{\partial{\cal E}_i}{\partial X}+\frac{1}{2}H
\sum_{i=2}^5\frac{\partial{\cal E}_i}{\partial H},\label{comp2}
\\
\Theta &=&-\frac{1}{6}\sum_{i=2}^5\frac{\partial{\cal E}_i}{\partial H},\label{comp3}
\end{eqnarray}
which compactify the above lengthy expressions.

Varying the action~(\ref{scalaraction1}) with respect to $\alpha$ and $\beta$,
we obtain the constraint equations
\begin{eqnarray}
\Sigma\alpha -\Theta\frac{\Vec{\nabla}^2}{a^2}\beta+3\Theta\dot\zeta
-{\cal G}_T\frac{\Vec{\nabla}^2}{a^2}\zeta&=&0,
\\
\Theta\alpha-{\cal G}_T\dot\zeta&=&0.
\end{eqnarray}
Using the constraint equations, we eliminate $\alpha$ and $\beta$
from the action~(\ref{scalaraction1}) and
finally arrive at
\begin{eqnarray}
S^{(2)}_S=\int\D t\D^3 x\,a^3\left[
{\cal G}_S
\dot\zeta^2
-\frac{{\cal F}_S}{a^2}
(\Vec{\nabla}\zeta)^2
\right]\label{scalar2},
\end{eqnarray}
where
\begin{eqnarray}
{\cal F}_S&:=&\frac{1}{a}\frac{\D}{\D t}\left(\frac{a}{\Theta}{\cal G}_T^2\right)
-{\cal F}_T,
\\
{\cal G}_S&:=&\frac{\Sigma }{\Theta^2}{\cal G}_T^2+3{\cal G}_T.
\end{eqnarray}

The analysis of the curvature perturbation hereafter is
completely parallel to that of the tensor perturbation.
The squared sound speed
is given by $c_S^2={\cal F}_S/{\cal G}_S$,
and ghost and gradient instabilities are avoided as long as
\begin{eqnarray}
{\cal F}_S>0\quad {\cal G}_S>0.\label{stsc}
\end{eqnarray}

In the case of k-inflation where $G_3=0=G_5$ and $G_4=\mpl^2/2$,
we have ${\cal F}_S=\mpl^2 \epsilon$.
This implies that the interesting regime $\dot H >0$ is prohibited
by the stability requirement in k-inflation~\cite{perkinf}.
However, the sign of $\dot H$ and the stability criteria are not correlated in more general situations.
This point was already clear in G-inflation and kinetic gravity
braiding for which $G_3\neq 0$~\cite{G-inf, KGB}.
Stable cosmological solutions with $\dot H>0$
offer a radical and
very interesting scenario of the earliest Universe~\cite{NV, genesis}.

The stability conditions~(\ref{stsc}) for the scalar perturbation
as well as~(\ref{stts}) for the tensor perturbation
have been derived in the case of the covariant Galileon,
for which $K=-c_2X, G_3=-c_3X/M^3, G_4=\mpl^2/2-c_4X^2/M^6$,
and $G_5=3c_5X^2/M^9$~\cite{tsujikaw}.
It can be verified that our general formulas
correctly reproduce the result of~\cite{tsujikaw}.

Using the new variables
\begin{eqnarray}
\D y_S:=\frac{c_S}{a}\D t,
\;\;
z_S:=\sqrt{2}a\left({\cal F}_S{\cal G}_S\right)^{1/4},
\;\;
u:=z_S\zeta,
\end{eqnarray}
the curvature perturbation is canonically normalized
and the action is now given by
\begin{eqnarray}
S_S^{(2)}=\frac{1}{2}\int\D y_S\D^3x\left[(u')^2-(\Vec{\nabla} u)^2+\frac{z_S''}{z_S}u^2\right],
\end{eqnarray}
where a prime denotes differentiation with respect to $y_S$.
Each perturbation mode exits the sound horizon when $k^2=z_S''/z_S\sim 1/y^2_S$,
where $k$ is the Fourier wavenumber.

The two independent solutions on superhorizon scales are
\begin{eqnarray}
\zeta ={\rm const}\quad{\rm and}\quad
\int^t\frac{\D t'}{a^3{\cal G}_S}.
\end{eqnarray}
During inflation, it may be assumed that ${\cal G}_S$ is slowly varying.
In this case, the second solution decays rapidly.
Note, however, that the non-trivial dynamics of the scalar field
can induce a temporal rapid evolution of ${\cal G}_S$,
which would affect the superhorizon behavior of the curvature perturbation
through the contamination of the second mode in the same way as in Ref.~\citen{enhance}.
Given the specific background dynamics, one can evaluate such an effect
using our general formulas.

Closely following the procedure we did in the case of the tensor
perturbation, we now evaluate the power spectrum of the primordial
curvature perturbation.  To do so, we assume that
$\epsilon\simeq\;$const,\footnote{By defining the variation parameter of the sound
velocity of scalar perturbations as $s_{S}:=\dot{c}_S/(H c_S) =
\left(f_{S} - g_{S} \right)/2$, the formulae with $f_S$ and/or $g_S$ can
be rewritten in terms of $c_S$.}
\begin{eqnarray}
f_S:=\frac{\dot{\cal F}_S}{H{\cal F}_S}\simeq{\rm const},\quad
g_S:=\frac{\dot{\cal G}_S}{H{\cal G}_S}\simeq{\rm const},
\end{eqnarray}
and then define
\begin{eqnarray}
\nu_S:=\frac{3-\epsilon+g_S}{2-2\epsilon-f_S+g_S}.
\end{eqnarray}
The power spectrum is given by
\begin{eqnarray}
{\cal P}_\zeta =\frac{\gamma_S}{2}\left.
\frac{{\cal G}_S^{1/2}}{{\cal F}_S^{3/2}}\frac{H^2}{4\pi^2}\right|_{-ky_S=1},
\end{eqnarray}
where $\gamma_S=2^{2\nu_S -3}|\Gamma(\nu_S)/\Gamma(3/2)|^2(1-\epsilon -f_S/2+g_S/2)$.
The spectral index is
\begin{eqnarray}
n_s-1=3-2\nu_S.
\end{eqnarray}
An exactly scale-invariant spectrum is obtained if
\begin{eqnarray}
\epsilon+\frac{3}{4}f_S-\frac{1}{4}g_S = 0.
\end{eqnarray}
Here again, $\epsilon$, $f_S$, and $g_S$ are not necessarily very small
(as long as $n_s-1\simeq 0$).

Taking now the limit $\epsilon, f_T, g_T, f_S, g_S \ll 1$,
the tensor-to-scalar ratio is given by
\begin{eqnarray}
r=16\left(\frac{{\cal F}_S}{{\cal F}_T}\right)^{3/2}
\left(\frac{{\cal G}_S}{{\cal G}_T}\right)^{-1/2}
=16\frac{{\cal F}_S}{{\cal F}_T}\frac{c_S}{c_T}.
\end{eqnarray}
Note that even in the de Sitter limit where $\epsilon, f_T, g_T, f_S, g_S \to 0$,
the scalar perturbation can be produced in general, $r\neq 0$.

In the case of potential-driven slow-roll inflation in \S\ref{slowroll},
we have, to leading order in slow-roll,
\begin{eqnarray}
{\cal F}_S&\simeq&\frac{X}{H^2}\left(\calk + 6H^2h_4\right)
+\frac{4\dot\phi X}{H}\left(h_3+H^2h_5\right),\label{slowf}
\\
{\cal G}_S&\simeq&\frac{X}{H^2}\left(\calk + 6H^2h_4\right)
+\frac{6\dot\phi X}{H}\left(h_3+H^2h_5\right),\label{slowg}
\end{eqnarray}
and ${\cal F}_T\simeq{\cal G}_T\simeq 2g_4$, where we used the slow-roll
equation $2g_4\epsilon+\dot{g}_4/H\simeq \dot\phi J/2H^2$.
In this case, we have
$c_T^2 \simeq 1$ and $n_T \simeq -(2 \epsilon +g_T)$ with $f_T \simeq
g_T \simeq \dot{g}_4/(Hg_4)$. If the
$\calk$ or $h_4$ term dominates in $J$, we have $c_S^2\simeq 1$ and
${\cal F}_S \simeq {\cal G}_S \simeq J\dot{\phi}/(2H^2) \simeq
g_4(2\epsilon+g_T)$, which yields the standard consistency relation:
\begin{equation}
  r \simeq -8 n_T.
\end{equation}
On the other hand, if the $h_3$ or $h_5$ term dominates, then we have
$c_S^2\simeq 2/3$, ${\cal F}_S \simeq 2J\dot{\phi}/(3H^2) \simeq
(4 g_4/3)(2\epsilon+g_T)$, and ${\cal G}_S \simeq J\dot{\phi}/H^2 \simeq
2 g_4(2\epsilon+g_T)$, which yields a new consistency relation\cite{kamada}:
\begin{equation}
  r \simeq -\frac{32\sqrt{6}}{9} n_T.
\end{equation}
Thus, we can discriminate which term dominates in the dynamics using
the consistency relations.

\section{Summary}

In this paper, generic inflation models
driven by a single scalar field have been studied.
Our gravity $+$ scalar-field system is
described by the generalized Galileons,
which do not give rise to higher derivatives in the field equations
despite the non-minimal coupling, {\it e.g.}, of the form $G(\phi, X)R$.
The class of inflation models 
is the most general ever proposed in the context of single-field inflation.

We have seen that
if the Lagrangian has a shift symmetry, $\phi\to\phi +c$,
de Sitter attractors are present and, hence, inflation can be driven by
$\phi$'s kinetic energy.
Reheating after kinetically driven inflation
is possible by breaking the shift symmetry, but the way
to break it depends on the explicit construction of the
originally shift-symmetric Lagrangian itself.

We have also derived slow-roll equations of motion
for potential-driven inflation,
in which the scalar-field dynamics is modified
by the higher-order Galileon terms.

We have determined the most generic quadratic actions for
tensor and scalar cosmological perturbations.
Using them, we have presented the stability criteria
for both types of perturbations.
The primordial power spectra have also been computed.
Note that, since the propagation speeds of the two types
of fluctuations can be different, we must evaluate the
power spectra for the same comoving wavenumber at different
epochs, which may have some consequence~\cite{Ringeval}.

It would be interesting to extend the present linear analysis
of the curvature perturbation
to non-linear order along the line of Refs.~\citen{Naruko, gradex}.
In relation to cosmological perturbations beyond linear order,
it would be important to evaluate
primordial non-Gaussianities generated from Generalized
G-inflation.

{\bf Acknowledgements}
This work was
supported in part by JSPS Grant-in-Aid for Research Activity Start-up
No. 22840011 (T.K.), Grant-in-Aid for Scientific Research
Nos. 23340058 (J.Y.) and 21740187 (M.Y.), and Grant-in-Aid for
Scientific Research on Innovative Areas No. 21111006 (J.Y.).

\appendix

\section{The Horndeski action, generalized Galileons, and non-minimal coupling to the Gauss-Bonnet term}

In 1974, Horndeski presented the most general action (in four dimensions)
constructed from the metric $g_{\mu\nu}$, the scalar field $\phi$, and their derivatives
$\partial g_{\mu\nu}, \partial^2 g_{\mu\nu}, \partial^3 g_{\mu\nu}$, $\cdots,
\partial\phi, \partial^2\phi, \partial^3\phi, \cdots $~\cite{Horndeski},
still having second-order field equations.
The Horndeski theory has been revisited recently by the authors of Ref.~\citen{Charmousis}.
In this appendix, we point out that the Horndeski theory and
the generalized Galileons are equivalent.

In terms of the notation of Ref.~\citen{Charmousis} but
using $X=-\partial_\mu\phi\partial^\mu\phi/2$ rather than $\rho=\partial_\mu\phi\partial^\mu\phi$,
the Lagrangian of the Horndeski theory is given by
\begin{eqnarray}
{\cal L}_H&=&\delta_{\mu\nu\sigma}^{\alpha\beta\gamma}
\biggl[
\kappa_1 \nabla^\mu\nabla_\alpha \phi R_{\beta \gamma}^{~~\;\nu\sigma}+\frac{2}{3}\kappa_{1X}
\nabla^\mu\nabla_\alpha\phi\nabla^\nu\nabla_\beta\phi\nabla^\sigma\nabla_\gamma\phi
+\kappa_3\nabla_\alpha\phi\nabla^\mu\phi R_{\beta\gamma}^{\;~~\nu\sigma}
\nonumber\\&&
+2\kappa_{3 X}\nabla_\alpha
\phi\nabla^\mu\phi\nabla^\nu\nabla_\beta\phi\nabla^\sigma\nabla_\gamma\phi
\biggr]
+\delta^{\alpha\beta}_{\mu\nu}\bigl[
(F+2W)R_{\alpha\beta}^{\;~~ \mu\nu}+2F_X\nabla^\mu\nabla_\alpha\phi\nabla^\nu\nabla_\beta\phi
\nonumber\\&&
+2\kappa_8\nabla_\alpha\phi\nabla^\mu\phi\nabla^\nu\nabla_\beta\phi\bigr]
-6\left(F_\phi+2W_\phi-X\kappa_8\right)\Box\phi+\kappa_9,
\label{LH}
\end{eqnarray}
where
$\delta^{\alpha_1\alpha_2...\alpha_n}_{\mu_1\mu_2...\mu_n}=n!\delta_{\mu_1}^{[\alpha_1}
\delta_{\mu_2}^{\alpha_2}...\delta_{\mu_n}^{\alpha_n]}$,
and $\kappa_1$, $\kappa_3$, $\kappa_8$, and $\kappa_9$ are arbitrary functions of $\phi$ and $X$.
We also have two functions $F=F(\phi,X)$ and $W=W(\phi)$,
and the former is constrained so that
$F_X=2(\kappa_3+2X\kappa_{3X}-\kappa_{1\phi})$, while
the latter can be absorbed into a redefinition of the former.
We are therefore left with four arbitrary functions of $\phi$ and $X$,
in accordance with the generalized Galileon.

The above Lagrangian can be
mapped to that of the generalized Galileon by identifying
\begin{eqnarray}
K&=&\kappa_9+4X\int^X\D X'\left(\kappa_{8\phi}-2\kappa_{3\phi\phi}\right),
\\
G_3&=&
6F_\phi-2X\kappa_8-8X\kappa_{3\phi}+2\int^X\D X'(\kappa_8-2\kappa_{3\phi}),
\\
G_4&=&2F-4X\kappa_3,
\\
G_5&=&-4\kappa_1,
\end{eqnarray}
where we redefined $F$ so that $F+2W \to F$.
Now we see that the two theories are in fact equivalent.
In deriving the Lagrangian~(\ref{LH}), Horndeski started
from the assumptions that are weaker than those made by Deffayet et
al.~\cite{GeneralizedGalileon},
although the latter worked in arbitrary dimensions.

Since the generalized Galileon is the most general theory in four dimensions
composed of $g_{\mu\nu}$, $\phi$, and their derivatives,
which gives the second-order field equations,
it must reproduce the non-minimal coupling to the Gauss-Bonnet term~\cite{g-b},
\begin{eqnarray}
\xi(\phi)\left(R^2-4R_{\mu\nu}R^{\mu\nu}
+R_{\mu\nu\rho\sigma}R^{\mu\nu\rho\sigma}\right),\label{gb}
\end{eqnarray}
which seems non-trivial at first glance.
One can show that, by taking
\begin{eqnarray}
K&=&8\xi^{(4)}X^2\left(3-\ln X\right),\label{K-GB}\\
G_3&=&4\xi^{(3)}X\left(7-3\ln X\right),\\
G_4&=&4\xi^{(2)}X\left(2-\ln X\right),\\
G_5&=&-4\xi^{(1)} \ln X,\label{G5-GB}
\end{eqnarray}
or, equivalently, $\kappa_1=\xi^{(1)}\ln X$, $\kappa_3=\xi^{(2)}\ln X$,
$\kappa_8=0$, and $\kappa_9=16\xi^{(4)}X^2$,
where $\xi^{(n)}:=\partial^n\xi/\partial\phi^n$,
the generalized Galileon indeed
reproduces the non-minimal coupling of the form~(\ref{gb}).
Probably the shortest way to confirm this fact is
to substitute $\kappa_i$ to the field equations presented in Ref.~\citen{Horndeski}
and to compare them with those obtained from~(\ref{gb}).

Similarly to $f(R)$ gravity,
the gravitational theory described by
\begin{eqnarray}
{\cal L}=\frac{R}{2}+f(\mathscr{G}),
\quad
\mathscr{G}:=R^2-4R_{\mu\nu}R^{\mu\nu}
+R_{\mu\nu\rho\sigma}R^{\mu\nu\rho\sigma},\label{fGB}
\end{eqnarray}
where $f(\mathscr{G})$ is an arbitrary function of
the Gauss-Bonnet term, contains an extra scalar degree of freedom,
and hence, (\ref{fGB}) must be recast in
the Lagrangian of the generalized Galileon. Noting that
the Lagrangian~(\ref{fGB}) can be equivalently written as
\begin{eqnarray}
{\cal L} =\frac{R}{2}+f(\phi)+f_\phi\left(\mathscr{G}-\phi\right),
\end{eqnarray}
and the non-minimal coupling $f_\phi\mathscr{G}$
is reproduced by Eqs.~(\ref{K-GB})--(\ref{G5-GB}),
it is now straightforward to translate~(\ref{fGB}) to the
generalized Galileon.

It is easy to see explicitly in the cosmological equations of motion that
the contribution from the non-minimal coupling~(\ref{gb}) can indeed be reproduced from
the non-trivial functions~(\ref{K-GB})--(\ref{G5-GB}).
Both the generalized Galileon with~(\ref{K-GB})--(\ref{G5-GB}) and
the Lagrangian~(\ref{gb}) give the following identical contributions
to the background and perturbation equations:
for the background gravitational field equations,
\begin{eqnarray}
{\cal E}&\supset&-24H^3\dot\xi,
\\
{\cal P}&\supset&8\left[H^2\ddot\xi+2H\left(H^2+\dot H\right)\dot\xi\right],
\end{eqnarray}
for the background equation of motion for $\phi$,
\begin{eqnarray}
P_\phi\supset 24H^2\left(\dot H+H^2\right)\xi_\phi,
\end{eqnarray}
and
for the quadratic actions of the tensor and scalar perturbations,
\begin{eqnarray}
{\cal F}_T\supset
8 \ddot\xi,
\quad
{\cal G}_T \supset 8H\dot\xi,
\quad
\Sigma \supset  -48H^3\dot\xi,
\quad
\Theta \supset 12H^2\dot\xi.
\end{eqnarray}

\section{Field equations}

In this Appendix, we present both gravitational- and scalar-field equations
derived from the action~(\ref{5}) for completeness.\footnote{In the
course of the preparation of this manuscript, 
we became aware that Gao has also calculated
gravitational field equations in the present model~\cite{Gao}.
After some iterations, his result has converged with ours
and we are in full agreement.}
Varying the action, we obtain 
\begin{eqnarray}
\delta\left(\sqrt{-g}\sum_{i=2}^5{\cal L}_i\right)
&=&\sqrt{-g}\left[\sum_{i=2}^5{\cal G}^i_{\mu\nu}\delta g^{\mu\nu}
+\sum_{i=2}^5\left(P_\phi^i-\nabla^\mu J_\mu^i\right)\delta\phi
\right]
\nonumber\\&&
+{\rm total~derivative},
\end{eqnarray}
where
\begin{eqnarray}
{\cal G}_{\mu \nu}^2&=&-\frac{1}{2}K_X\nabla_\mu\phi\nabla_\nu\phi-\frac{1}{2}Kg_{\mu\nu},
\\
{\cal G}_{\mu \nu}^3&=&\frac{1}{2}G_{3X}\Box\phi\nabla_\mu\phi\nabla_\nu\phi
+\nabla_{(\mu}G_3\nabla_{\nu)}\phi-\frac{1}{2}g_{\mu\nu}\nabla_\lambda G_3\nabla^\lambda\phi ,
\\
{\cal G}_{\mu \nu}^4&=&
G_4G_{\mu\nu}-\frac{1}{2}G_{4X}R\nabla_\mu\phi\nabla_\nu\phi
-\frac{1}{2}G_{4XX}\left[(\Box\phi)^2-(\nabla_\alpha\nabla_\beta\phi)^2\right]\nabla_\mu
\phi\nabla_\nu\phi
\cr&&
-G_{4X}\Box\phi\nabla_\mu\nabla_\nu\phi
+G_{4X}\nabla_\lambda\nabla_\mu\phi\nabla^\lambda\nabla_\nu\phi
+2\nabla_\lambda G_{4X}\nabla^\lambda\nabla_{(\mu}\phi\nabla_{\nu)}\phi
\cr&&
-\nabla_\lambda G_{4X}\nabla^\lambda\phi\nabla_\mu\nabla_\nu\phi
+g_{\mu\nu}\left(G_{4\phi}\Box\phi-2XG_{4\phi\phi}\right)
\cr&&
+g_{\mu\nu}\biggl\{-2G_{4\phi X}\nabla_\alpha\nabla_\beta\phi\nabla^\alpha\phi\nabla^\beta\phi
+G_{4XX}\nabla_\alpha\nabla_\lambda\phi\nabla_\beta\nabla^\lambda\phi\nabla^\alpha\phi\nabla^\beta\phi
\cr&&
+\frac{1}{2}G_{4X}\left[(\Box\phi)^2-(\nabla_\alpha\nabla_\beta\phi)^2\right]\biggr\}
+2\Bigl[G_{4X}R_{\lambda(\mu}\nabla_{\nu)}\phi\nabla^\lambda\phi
\cr&&
-\nabla_{(\mu}G_{4X}\nabla_{\nu)}\phi\Box\phi\Bigr]
-g_{\mu\nu}\left[
G_{4X}R^{\alpha\beta}\nabla_\alpha\phi\nabla_\beta\phi-\nabla_\lambda G_{4X}\nabla^\lambda\phi
\Box\phi\right]
\cr&& +G_{4X}R_{\mu\alpha\nu\beta}\nabla^\alpha\phi\nabla^\beta\phi
-G_{4\phi}\nabla_\mu\nabla_\nu\phi-G_{4\phi\phi}\nabla_\mu\phi\nabla_\nu\phi
\cr&&
+2G_{4\phi X}\nabla^\lambda\phi\nabla_\lambda
\nabla_{(\mu}\phi\nabla_{\nu)}\phi
-G_{4XX}\nabla^\alpha\phi\nabla_\alpha\nabla_\mu\phi
\nabla^\beta\phi\nabla_\beta\nabla_\nu\phi,
\\
{\cal G}_{\mu \nu}^5&=&
G_{5X}R_{\alpha\beta}\nabla^\alpha\phi\nabla^\beta\nabla_{(\mu}\phi
\nabla_{\nu)}\phi
-G_{5X}R_{\alpha(\mu}\nabla_{\nu)}\phi \nabla^\alpha\phi\Box\phi
\cr&&
-\frac{1}{2}G_{5X}R_{\alpha\beta}\nabla^\alpha\phi\nabla^\beta\phi\nabla_\mu\nabla_\nu\phi
-\frac{1}{2}G_{5X}R_{\mu\alpha\nu\beta}\nabla^\alpha\phi\nabla^\beta\phi\Box\phi
\cr&&
+G_{5X}R_{\alpha\lambda\beta(\mu}\nabla_{\nu)}\phi\nabla^\lambda\phi
\nabla^\alpha\nabla^\beta\phi
+G_{5X}R_{\alpha\lambda\beta(\mu}\nabla_{\nu)}\nabla^\lambda\phi\nabla^\alpha\phi\nabla^\beta\phi
\cr&&
-\frac{1}{2}\nabla_{(\mu}\left[G_{5X}\nabla^\alpha\phi\right]\nabla_\alpha\nabla_{\nu)}\phi\Box\phi
+\frac{1}{2}\nabla_{(\mu}\left[G_{5\phi}\nabla_{\nu)}\phi\right]\Box\phi
\cr&&
-\nabla_\lambda\left[G_{5\phi}\nabla_{(\mu}\phi\right]\nabla_{\nu)}\nabla^\lambda\phi
\cr&&
+\frac{1}{2}\left[
\nabla_\lambda \left(G_{5\phi}\nabla^\lambda\phi\right)-\nabla_\alpha\left(G_{5X}\nabla_\beta\phi\right)
\nabla^\alpha\nabla^\beta\phi
\right]\nabla_\mu\nabla_\nu\phi
\cr&&
+\nabla^{\alpha}G_5\nabla^{\beta}\phi R_{\alpha(\mu\nu)\beta}
-\nabla_{(\mu}G_5G_{\nu)\lambda}\nabla^\lambda\phi
\cr&&
+\frac{1}{2}\nabla_{(\mu}G_{5X}\nabla_{\nu)}\phi\left[
(\Box\phi)^2-(\nabla_\alpha\nabla_\beta\phi)^2\right]
\cr&&
-\nabla^\lambda G_5R_{\lambda(\mu}\nabla_{\nu)}\phi
+\nabla_\alpha\left[G_{5X}\nabla_\beta\phi\right]\nabla^\alpha\nabla_{(\mu}\phi
\nabla^\beta\nabla_{\nu)}\phi
\cr&&
-\nabla_\beta G_{5X}\left[
\Box\phi\nabla^\beta\nabla_{(\mu}\phi-\nabla^\alpha\nabla^\beta\phi
\nabla_\alpha\nabla_{(\mu}\phi\right]\nabla_{\nu)}\phi
\cr&&
+\frac{1}{2}\nabla^\alpha\phi\nabla_\alpha G_{5X}\left[
\Box\phi\nabla_\mu\nabla_\nu\phi -\nabla_\beta\nabla_\mu\phi\nabla^\beta\nabla_\nu\phi\right]
\cr&&
-\frac{1}{2} G_{5X} G_{\alpha\beta}\nabla^\alpha\nabla^\beta\phi\nabla_\mu\phi\nabla_\nu\phi
-\frac{1}{2}G_{5X}\Box\phi\nabla_\alpha\nabla_\mu\phi\nabla^\alpha\nabla_\nu\phi
\cr&&
+\frac{1}{2}G_{5X}(\Box\phi)^2\nabla_\mu\nabla_\nu\phi
+\frac{1}{12}G_{5XX}\bigl[(\Box\phi)^3-3\Box\phi(\nabla_\alpha\nabla_\beta\phi)^2
\cr&&
+2(\nabla_\alpha\nabla_\beta\phi)^3\bigr]\nabla_\mu\phi\nabla_\nu\phi
+\frac{1}{2}\nabla_\lambda G_5 G_{\mu\nu}\nabla^\lambda\phi
\cr&&
+g_{\mu\nu}\Biggl\{
-\frac{1}{6}G_{5X}\left[(\Box\phi)^3-3\Box\phi(\nabla_\alpha\nabla_\beta\phi)^2
+2(\nabla_\alpha\nabla_\beta\phi)^3\right]
+\nabla_\alpha G_5R^{\alpha\beta}\nabla_\beta\phi
\cr&&
-\frac{1}{2}\nabla_\alpha\left(G_{5\phi}\nabla^\alpha\phi\right)\Box\phi
+\frac{1}{2}\nabla_\alpha\left(G_{5\phi}\nabla_\beta\phi\right)\nabla^\alpha\nabla^\beta\phi
-\frac{1}{2}\nabla_\alpha G_{5X}\nabla^\alpha X\Box\phi
\cr&&
+\frac{1}{2}
\nabla_\alpha G_{5X}\nabla_\beta X\nabla^\alpha\nabla^\beta\phi
-\frac{1}{4}\nabla^\lambda G_{5X}\nabla_\lambda\phi\left[
(\Box\phi)^2-(\nabla_\alpha\nabla_\beta\phi)^2\right]
\cr&&
+\frac{1}{2}G_{5X}R_{\alpha\beta}\nabla^\alpha\phi\nabla^\beta\phi \Box\phi
-\frac{1}{2}G_{5X}R_{\alpha\lambda\beta\rho}
\nabla^\alpha\nabla^\beta\phi\nabla^\lambda\phi\nabla^\rho\phi
\Biggr\},
\end{eqnarray}
\begin{eqnarray}
P_\phi^2&=&K_\phi,
\\
P_\phi^3&=&\nabla_\mu G_{3\phi}\nabla^\mu\phi,
\\
P_\phi^4&=&G_{4\phi}R+G_{4\phi X}\left[(\Box\phi)^2-(\nabla_\mu\nabla_\nu\phi)^2\right],
\\
P_\phi^5&=&-\nabla_\mu G_{5\phi }G^{\mu\nu} \nabla_\nu\phi -
\frac{1}{6}G_{5\phi X}
\left[
(\Box\phi)^3-3\Box\phi (\nabla_\mu\nabla_\nu \phi)^2+2(\nabla_\mu\nabla_\nu \phi)^3
\right],
\end{eqnarray}
and
\begin{eqnarray}
J_\mu^2&=&-{\cal L}_{2X}\nabla_\mu\phi,
\\
J_\mu^3&=&-{\cal L}_{3X}
\nabla_\mu\phi+ G_{3X}\nabla_\mu X+2G_{3\phi}\nabla_\mu\phi,
\\
J_\mu^4&=&-{\cal L}_{4X}\nabla_\mu\phi+
2G_{4X}R_{\mu\nu}\nabla^\nu\phi
-2G_{4XX}\left(\Box\phi\nabla_\mu X-\nabla^\nu X\nabla_\mu\nabla_\nu\phi\right)
\cr&&
-2G_{4\phi X}\left(\Box\phi\nabla_\mu \phi +\nabla_\mu X\right),
\\
J_\mu^5&=&
-{\cal L}_{5X}\nabla_\mu\phi-2G_{5\phi }G_{\mu\nu}\nabla^\nu\phi
\cr&&
-G_{5X}\left[G_{\mu\nu}\nabla^\nu X+
R_{\mu\nu}\Box\phi \nabla^\nu\phi-R_{\nu\lambda}\nabla^\nu\phi\nabla^\lambda\nabla_\mu\phi
-R_{\alpha\mu\beta\nu}\nabla^\nu\phi\nabla^\alpha\nabla^\beta\phi
\right]
\cr
&&
+G_{5XX}\left\{
\frac{1}{2}\nabla_\mu X\left[(\Box\phi)^2-(\nabla_\alpha\nabla_\beta\phi)^2\right]
-\nabla_\nu X\left(\Box\phi \nabla_\mu\nabla^\nu\phi-\nabla_\alpha\nabla_\mu\phi
\nabla^\alpha\nabla^\nu\phi\right)
\right\}
\cr
&&
+G_{5\phi X}\left\{
\frac{1}{2}\nabla_\mu\phi\left[(\Box\phi)^2-(\nabla_\alpha\nabla_\beta\phi)^2\right]
+\Box\phi\nabla_\mu X-\nabla^\nu X\nabla_\nu\nabla_\mu \phi
\right\}.
\end{eqnarray}
The gravitational- and scalar-field equations are thus given by
\begin{eqnarray}
\sum_{i=2}^5{\cal G}_{\mu\nu}^i=0,
\quad \nabla^\mu\left(
\sum_{i=2}^5 J_\mu^i \right)= \sum_{i=2}^5 P^i_\phi.
\end{eqnarray}
One might worry that $\nabla^\mu J_\mu^i$ gives rise to higher derivatives as
$J_\mu^i$ apparently contains second-order derivatives.
However, this is not the case because the commutations of higher derivatives
can be replaced by the curvature tensors and thus are canceled. For instance, one has
$\nabla_\mu\left(\Box\phi\nabla^\mu+\nabla^\mu X\right)=
(\Box\phi)^2-(\nabla_\alpha\nabla_\beta\phi)^2-R_{\mu\nu}\nabla^\mu\phi\nabla^\nu\phi$.


\end{document}